# Study of Cosmic-Ray Modulation during the Recent Unusual Minimum and Mini Maximum of Solar Cycle 24

**O. P. M. Aslam · Badruddin**

*Department of Physics, Aligarh Muslim University, Aligarh-202002, India.*

**Abstract** After a prolonged and deep solar minimum at the end of Cycle 23, the current Solar Cycle 24 is one of the lowest cycles. These two periods of deep minimum and mini maximum are separated by a period of increasing solar activity. We study the cosmic-ray intensity variation in relation with the solar activity, heliospheric plasma and field parameters, including the heliospheric current sheet, during these three periods (phases) of different activity level and nature: (a) a deep minimum, (b) an increasing activity period and (c) a mini maximum. We use neutron monitor data from stations located around the globe to study the rigidity dependence on modulation during the two extremes, *i.e.*, minimum and maximum. We also study the time lag between the cosmic-ray intensity and various solar and interplanetary parameters separately during the three activity phases. We also analyze the role of various parameters, including the current sheet tilt, in modulating the cosmic-ray intensity during the three different phases. Their relative importance and the implications of our results are also discussed.

## 1. Introduction

Solar modulation of galactic cosmic rays (GCR) has been a subject of intense research, especially to assess the continuously changing behaviour of the sun and its influence on cosmic-rays. The modulation of GCR intensity associated with the 11-year solar activity cycle has been studied from the past several decades (*e.g.*, Forbush, 1954; Burlaga *et al.*, 1985; Venkatesan and Badruddin, 1990; Storini *et al.*, 1995; Sabbah and Rybansky, 2006; Kudela, 2009; Heber, 2013; Chowdhury, Kudela, and Dwivedi, 2013; and references therein). The long-term GCR modulation shows an 22-year cycle related to the solar magnetic cycle as the solar polarity reverses near solar maximum of every activity cycle. The polarity dependent effects on cosmic rays have also been an area of active research (*e.g.*, Jokipii, Levy, and Hubbard, 1977; Potgieter and Moraal, 1985; Smith and Thomas, 1986; Cliver and Ling, 2001; Kota 2013; Potgieter, 2014; Laurenza *et al.*, 2014; Potgieter *et al.*, 2014; Thomas, Owens, and Lockwood, 2014; Thomas *et al.*, 2014; and references therein).

The transport of GCR particles in the heliosphere is subjected to four distinct types of modulation mechanisms: convection, adiabatic energy changes, gradient and curvature drifts, and diffusion (Kota, 2013). It has been an accepted paradigm that drifts are the dominant processes during low to moderate solar activity while diffusion is considered to dominate the cosmic-ray modulation during solar maximum (see also , Jokipii and Wibberenz, 1998; Paccini and Usoskin, 2014). However, it has been suggested that all the four basic mechanisms (convection, diffusion, adiabatic energy change and drifts) are important, but their relative importance varies during different phases (*e.g.*, increasing and decreasing activity, maximum and minimum) throughout the solar cycle (*e.g.*, McKibben *et al*., 1995; McDonald, Nand, and McGuire, 1998) and there are evidences (*e.g.*, Badruddin, Singh, and Singh, 2007; Singh, Singh, and Badruddin, 2008; Aslam and Badruddin, 2012, 2014) that indicate such variations. -It has been recently suggested (see Potgieter, 2014) that GCR modulation is an intriguing interplay among basically four mechanisms that changes over the solar cycle and from one solar cycle to another.

The 11– and 22 –year GCR intensity modulation in antiphase with solar activity shows some time lag. This time lag has been observed to be different in odd-numbered solar activity cycles (*e.g.*, Solar Cycles 21, 23) from even numbered Solar Cycles (*e.g.*, Solar Cycles 20, 22). The time lag is also differs in positive polarity epochs (when northern solar pole has positive polarity, *i.e.*, outward directed field) from negative solar polarity epochs (when the north solar pole has the negative polarity, *i.e.*, inward directed field) (*e.g.*, Mavromichalaki, Belehaki, and Rafois, 1998; Kane, 2003; Badruddin, Singh, and Singh, 2007; Singh, Singh, and Badruddin, 2008; Inceoglu *et al*., 2014; Kane 2014). However, a number of similarities and differences in GCR intensity modulation during different phases of the solar cycle still need to be understood (*e.g.*, Potgieter, 2014; Aslam and Badruddin, 2014; Thomas, Owens, and Lockwood, 2014).

The recent solar minimum between Solar Cycles 23 and 24 was quite unusual, longer and deeper, and the present maximum is the smallest since that of Solar Cycle 14. Compared to the other three minima during the space age, the recent minimum was unusual in many aspects (Jian, Russell, and Luhmann, 2011; Pacini and Usoskin, 2015). Some of them are: a record low interplanetary magnetic field intensity, slower solar wind, reduced solar wind dynamic pressure, and weaker solar polar magnetic fields. Cosmic-ray modulation during this peculiar minimum has been studied using ground-based neutron monitor data (*e.g.,* Heber *et al.,* 2009; Moraal and Stoker, 2010; Badruddin 2011; Aslam and Badruddin, 2012; Cliver, Richardson and Ling, 2013; Ahluwalia 2014; Pacini and Usoskin, 2015), as well as spacecraft data (*e.g.,* Mewaldt *et al.*, 2010; McDonald, Webber, and Reames, 2010; Leske *et al.*, 2013; Guo and Florinski, 2014; Strauss and Potgieter, 2014) and many peculiarities have been observed. Some of these are: (a) the GCR flux was the -highest of the space age, (b) the energy spectrum was softer than expected (c) the modulation was weaker than in three previous minima at low energy particles, that are recorded by high-latitude Earth-based neutron monitors at polar sites (see Aslam and Badruddin, 2012; Strauss and Potgieter, 2014; Pacini and Usoskin, 2015, and references therein).

In this work, we concentrate mainly on the two recent peculiar periods, recent deep minimum (DeepMin) between Solar Cycles 23 and 24 and the mini maximum (MiniMax) of Solar Cycle 24. Additionally, we consider the period of increasing solar activity between these two periods of special characteristics.

## 2. Results and Discussion

### 2.1 Time Variation of the GCR Intensity and Solar and Interplanetary Parameters

The Cosmic-ray flux has been monitored continually since the last several decades by ground based neutron monitors (NMs). The 27-day average cosmic-ray intensity recorded at the Oulu NM from April 1964 up to November 2014 is shown in the upper panel of Figure 1. The Sunspot number is usually considered as a key indicator of solar activity because of the length of the available record. To compare the temporal variation in cosmic-ray intensity with solar activity, the lower panel of Figure 1 shows the solar rotation averaged sunspot number variation for the corresponding periods during different solar activity cycles (Solar Cycle 20 to Solar Cycle 24) and different solar polarity epochs ($A>0$ and $A<0$), here $A$ represents the polarity of the solar magnetic field, which is taken as positive ($A>0$) when dominant polar field is outward in the northern and inward in the southern hemisphere, and it is taken as negative ($A<0$) if the opposite is the state of the solar polar field in northern and southern hemispheres. The period of interest for this study [DeepMin, ascending phase and MiniMax] are marked between vertical lines on the right side of this figure. More specifically, the starting of DeepMin (with a blue dashed line), ascending phase (with a red line) and MiniMax (with a dash-dotted line) are marked. All the three phases under consideration fall into the period of negative polarity ($A<0$) - of the solar magnetic field.

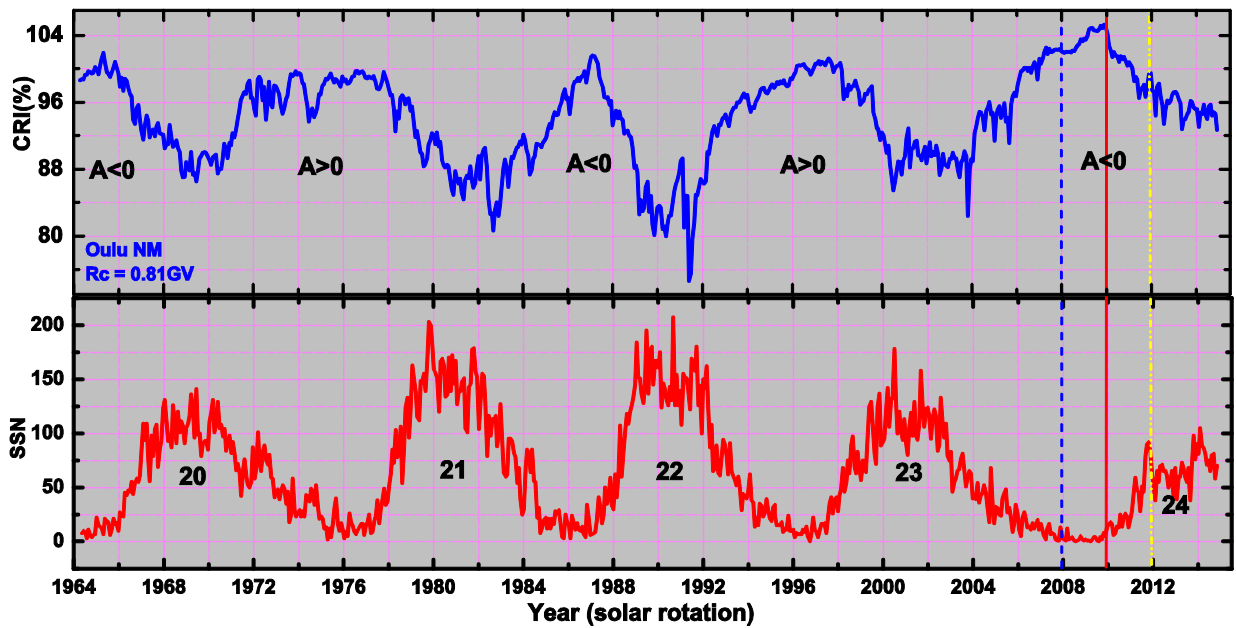

**Figure 1** The variation of relative cosmic-ray intensity and solar activity (SSN) with a 27-day average from April 1964 to November, 2014. The beginning of DeepMin, increasing and MiniMax phases are indicated by vertical lines.

These two periods are unique at least in the sense that a record high GCR intensity was recorded during DeepMin compared to intensities recorded during previous minima and the MiniMax is a solar maximum with lowest SSN among various solar maxima since the beginning of the space era.

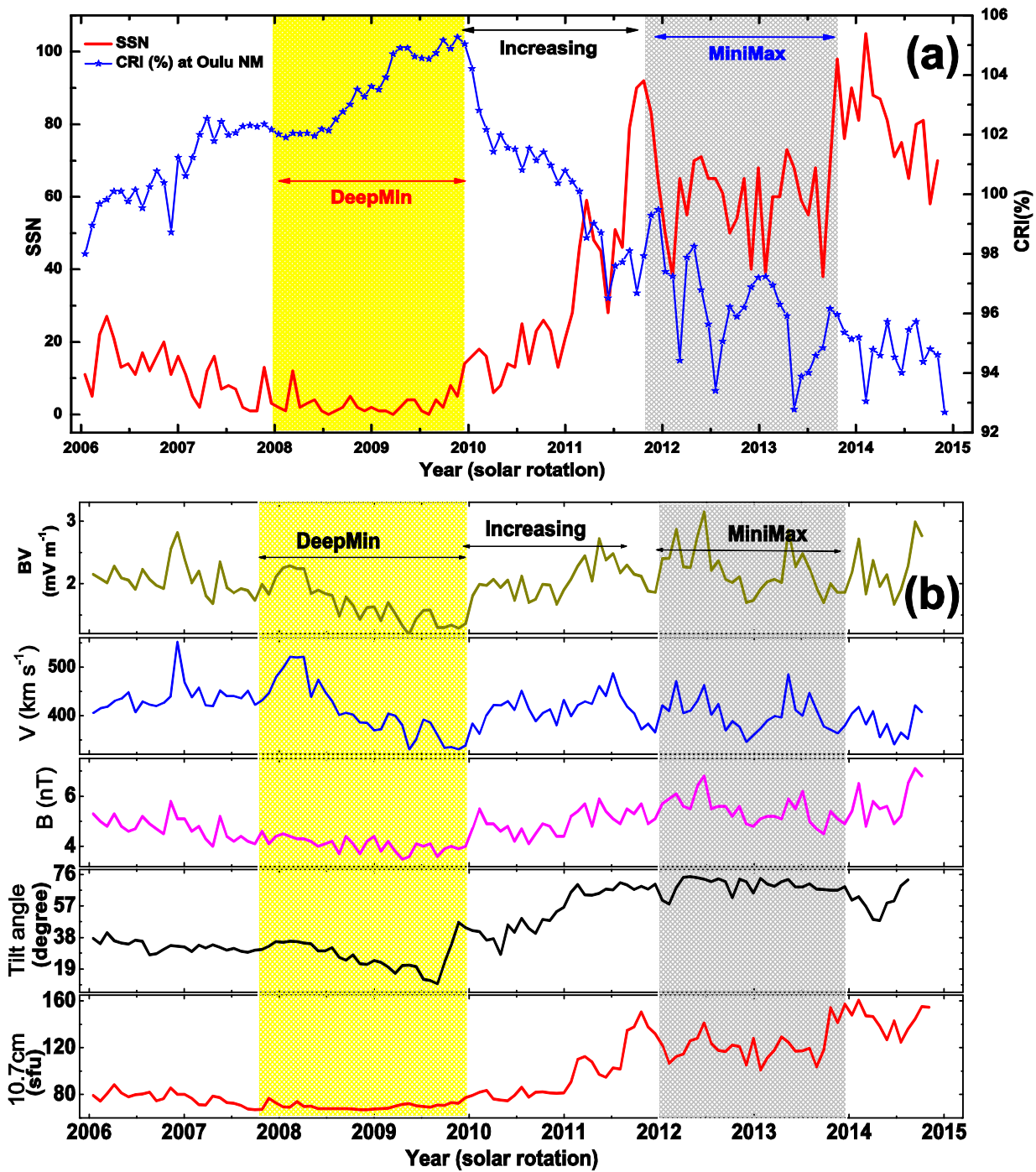

**Figure 2 (a)** (top panel) The variation of cosmic-ray intensity and solar activity (SSN) with a 27-day average from January 2006 to November 2014. The shaded portions indicating the DeepMin and MiniMax phases. (**b**) (bottom panel) Variation of the 10.7 cm solar radio flux, tilt of the HCS, IMF strength (*B*), solar wind velocity (*V*) and the product *BV* over a solar rotation average during the period January 2006 – November 2014. The shaded portions indicate the DeepMin and MiniMax phases.

In order to compare the three periods with variations in various parameters, we have plotted the GCR intensity and the SSN in Figure 2 (a) for the limited period 2006 – 2014. The variation of the solar wind parameters, the velocity *V* (km s$^{-1}$), the magnetic field *B* (nT) and the electric field *BV* (mV m$^-$ [1]) along with the tilt of the heliospheric current sheet (degree) and the 10.7 cm solar radio flux (sfu) are shown in Figure 2 (b), for the period 2006 – 2014 averaged over a solar rotation. The period of DeepMin and MiniMax are shaded. The 10.7cm solar radio-flux variation is similar to the SSN, *i.e.*, after the DeepMin of 2009 it reached a peak by the end of 2011 and, after a decrease, it maintained a lower level, although fluctuating, for a period of about 2 years. This period is marked as MiniMax and has a duration of 25 solar rotations (24 October 2011 to 29 August 2013). To compare the statistical analysis results of the three solar activity phases, we took the same duration for the three phases. Thus, 25 solar rotations just before the end of the deep solar minimum between Solar Cycles 23 and 24 as DeepMin (20 December 2007 to 25 October 2009) and 25 solar rotations after the deep solar minimum as ascending phase (21 November 2009 to 27 September 2011) were considered for the analysis. The tilt angle reached at its lowest value of 10.1 degree in DeepMin (October 2009), and then it started increasing. During the MiniMax period, the title angle is around 70 degree, which is almost similar to the tilt angle reached during the previous three maxima. The interplanetary magnetic field reached at its minimum level during the DeepMin between Solar Cycles 23 and 24 (3.5 nT averaged over a solar rotation), and it gradually started increasing along the last solar rotations of 2009. During the MiniMax period, the interplanetary magnetic field is around 5 nT, and the magnitude and fluctuations are less compared to the same phase in previous solar cycles (see Aslam and Badruddin 2014; Ahluwalia and Ygbuhay, 2015). The solar rotation averaged solar wind velocity reached a low value in 2009 (331 km s$^{-1}$), rose to a level of about 450 km s$^{-1}$, and afterwards it stayed at a relatively low value, fluctuating around  400 km s$^{-1}$ during MiniMax.

The upper panel of Figure 2 (b) shows the temporal variation of the interplanetary electric field ($E = BV$) averaged over a solar rotation for the period 2006  2014. The electric field reached a very low level in 2009 (1.2 mV m$^{-1}$), and gradually increased to  3 mV m$^{-1}$ during MiniMax, but this level is low when compared to similar phases of  previous solar cycles (see Aslam and Badruddin, 2014).

## 2.2 Time Lag between GCR Intensity and Solar and Interplanetary Parameters during DeepMin, MiniMax and the Ascending Phases of Solar Cycle 24

We can observe from Figure 2 that there appears a shift in the occurrence of peaks and dips in GCR intensity changes and some of the solar and interplanetary parameters due to a possible time lag between them. For determining the time lag between the GCR intensity and solar and interplanetary parameters, we have taken the three phases with equal duration (25 solar rotations,  2 years). We calculated the Pearson correlation coefficients between the GCR intensity and other parameters by introducing successively a time lag of one rotation from -5 to +30 solar rotations. The correlation coefficients so obtained are plotted in Figures 3 and 4. We can infer the most probable time lag between the GCR intensity and the individual parameters from the optimum values of the correlation coefficients. The left column of Figures 3 and 4 shows the variations in the time lag correlation coefficients of various solar and interplanetary parameters during the deep minimum between Solar Cycles 23 and 24, the central column shows the variations of correlation coefficients during the increasing phase of Solar Cycle 24 and the right column shows the variations of correlation coefficients during MiniMax phase.

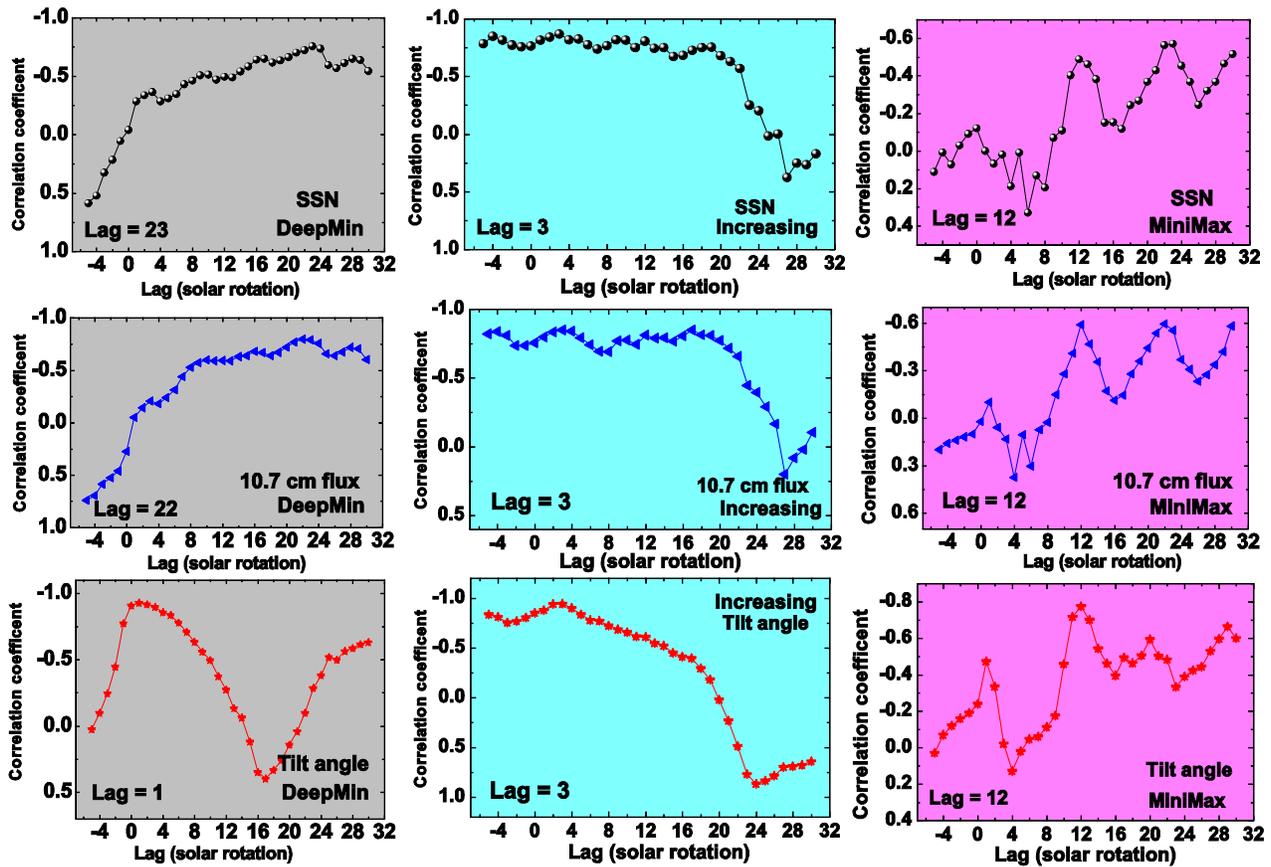

**Figure 3** The variation of time lag correlation between the 27-day averaged (**a**) GCR intensity and sunspot number (top three panels) , (**b**) GCR intensity and 10.7cm solar radio flux (middle three panels) and (**c**) GCR intensity and HCS tilt angle (three bottom panels) during DeepMin, increasing (middle column) and MiniMax phases.

We observe that the time lag is different for solar parameters (SSN and 10.7 cm solar radio flux) and the tilt angle (  ) during the three phases of solar activity, but for the solar wind parameters time lag is almost zero in all three phases of solar activity averaged over a solar rotation period. The zero time lag for solar wind parameters (interplanetary magnetic field IMF strength and solar wind velocity) is surprising as the solar wind magnetic field is known to play an important role in GCR intensity modulation. However, earlier studies (*e.g.,* Chowdhury, Kudela, and Dwivedi, 2013) have also reported a zero time lag between IMF and GCR during Solar Cycle 23. One possibility, as proposed by Chowdhury, Kudela, and Dwivedi (2013), is that the local disturbances such as propagating shocks, coronal mass ejections *etc.* dominate at 1 AU over the effects of distant merged interaction regions (MIRs) and Global Merged Interaction Regions (GMIRs) at large distances during this period. However, further analysis is required to clarify this. During the deep minimum the time lag profile of the HCS tilt is quite different as compared with the profile of SSN and 10.7cm solar radio flux, but during the increasing and MiniMax phases, all the three profiles show similar time-lag variation. As given in Table 1, the time lags for SSN and 10.7cm solar radio flux are 22 rotations during the deep minimum, 3 rotations during the increasing phase and 12 solar rotations during the MiniMax phase of Solar Cycle 24. In other words, the time lag between GCR intensity and solar parameters (SSN and 10.7 cm solar radio flux) is not the same during the three phases; it is shorter (3 solar rotations) during the increasing phase, intermediate (12 solar rotations) during MiniMax, and longer ( 22 solar rotations) during DeepMin phases.

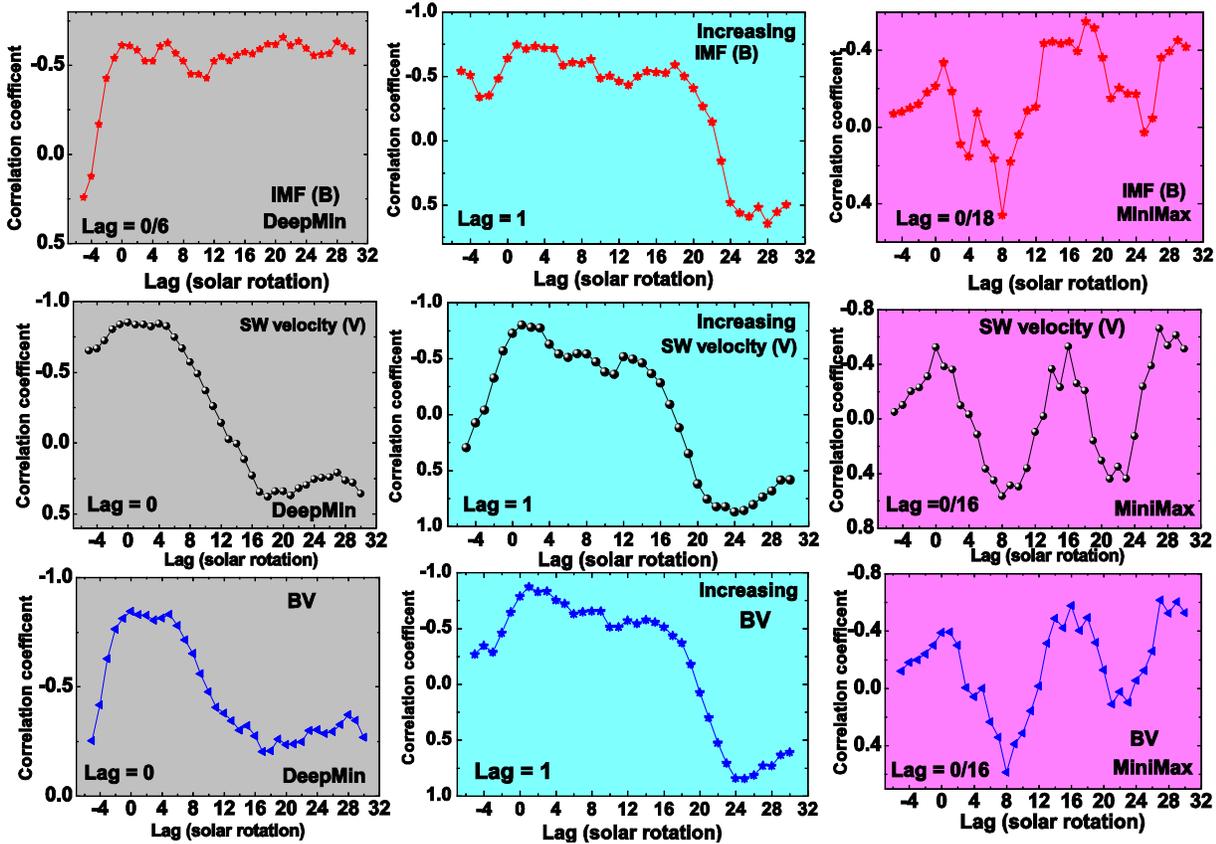

**Figure 4** The variation of time lag correlation between the 27-day averaged **(a)** GCR intensity and interplanetary magnetic field (*B*) (top three panels), **(b)** GCR intensity and solar wind velocity (*V*) (middle three panels) and **(c)** GCR intensity and *BV* (three bottom panels) during DeepMin, increasing (middle column) and MiniMax phases.

The time lag for the tilt angle during DeepMin is only one solar rotation against 22 solar rotations for SSN and 10.7 cm solar radio flux during the same period. The time lag is the same for tilt angle, SSN, and 10.7 cm solar radio flux during increasing (3 solar rotations) and MiniMax (12 solar rotations) phases. However, the lag between GCR intensity and solar wind parameters, solar wind velocity (*V*), IMF (*B*) and interplanetary electric field (*BV*) is found to be zero averaged over a solar rotation during all three phases, DeepMin, ascending and MiniMax. A peculiar observation during MiniMax is that the time lag correlation shows some sort of periodic behaviour from negative to positive correlation between GCR intensity and interplanetary parameters (*V, B,* and *BV*). The second peak with negative correlation coefficient after zero time appears at 16 – 18 solar rotations ( 1.3 year), a periodicity observed in the IMF and solar wind (Mursula and Vilppola, 2004) and several solar parameters (*e.g.*, see Cho, Hwang, and Park, 2014; Deng *et al*., 2015; and references therein).

### 2.2.1 *Observation of 27-day Recurrence in the Time Lag between GCR Intensity and Solar and Interplanetary Parameters during DeepMin*

In addition to study the time lag over the solar rotation average scale, we have also calculated the time lag correlation using daily data during DeepMin, increasing and MiniMax periods (Figures 5, 6 and 7).

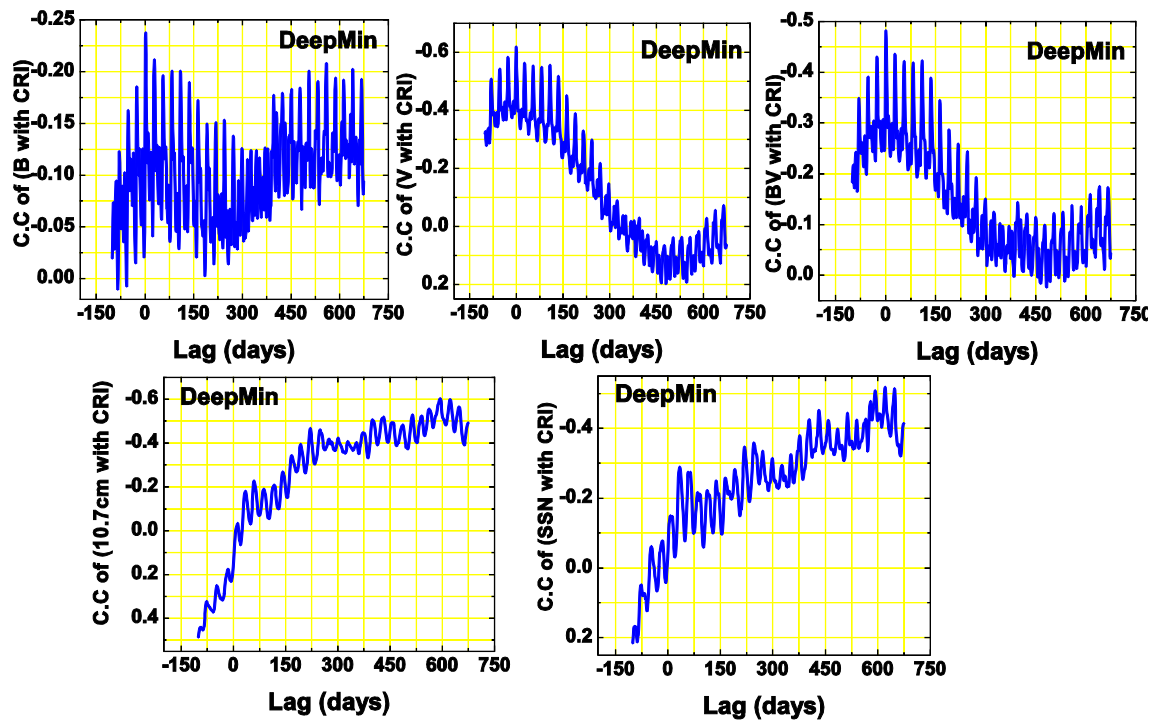

**Figure 5** The variation of time-lag correlation between the daily GCR intensity and solar and interplanetary parameters (SSN, 10.7cm solar radio flux, *V*, *B* and *BV*) during the DeepMin phase.

In addition to corroborating the general shape of the lag-correlation graph shown in Figures 3 and 4, we observe interesting periodicities in the correlation between GCR and various parameters (SSN, 10.7 cm solar radio flux, *V*, *B*, and *BV*) during DeepMin. The 27-day periodicity in time lag correlations of all five plots are shown in Figure 5. In addition to 27-day periodicity, two more peaks (at 9-days and 18-days) are seen in the top three plots of Figure 5. These may be harmonics of the 27-day periodicity. To illustrate this point further, we have plotted an expanded version of Figure 5 for limited days (see Figure 8).

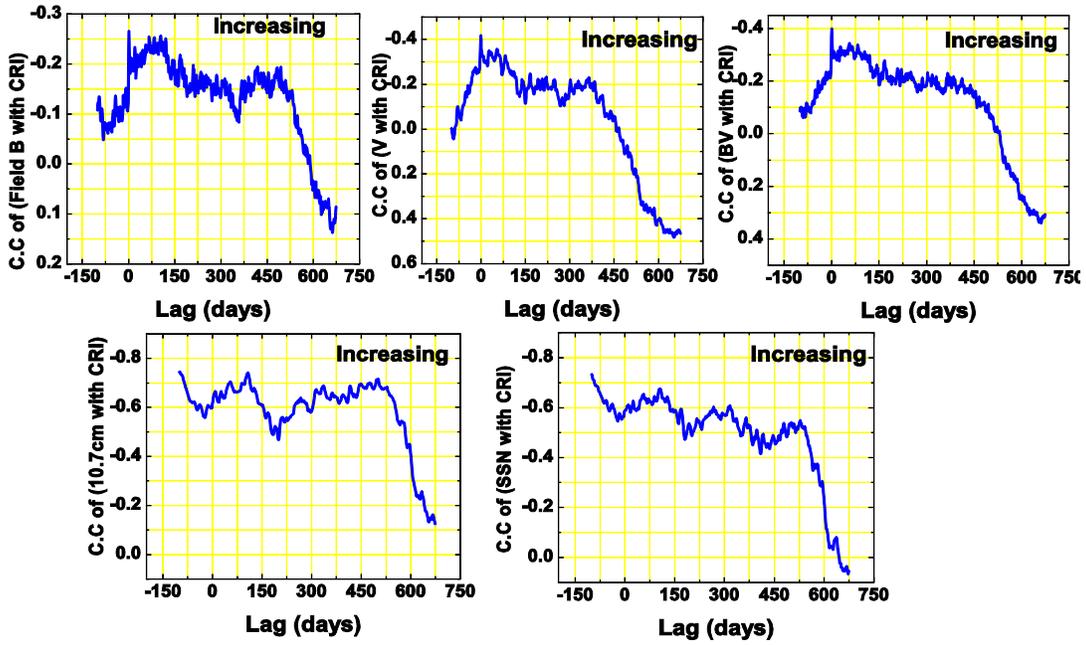

**Figure 6** The variation of time-lag correlation between the daily GCR intensity and solar and interplanetary parameters (SSN, 10.7cm solar radio flux, *V*, *B* and *BV*) during the increasing phase.

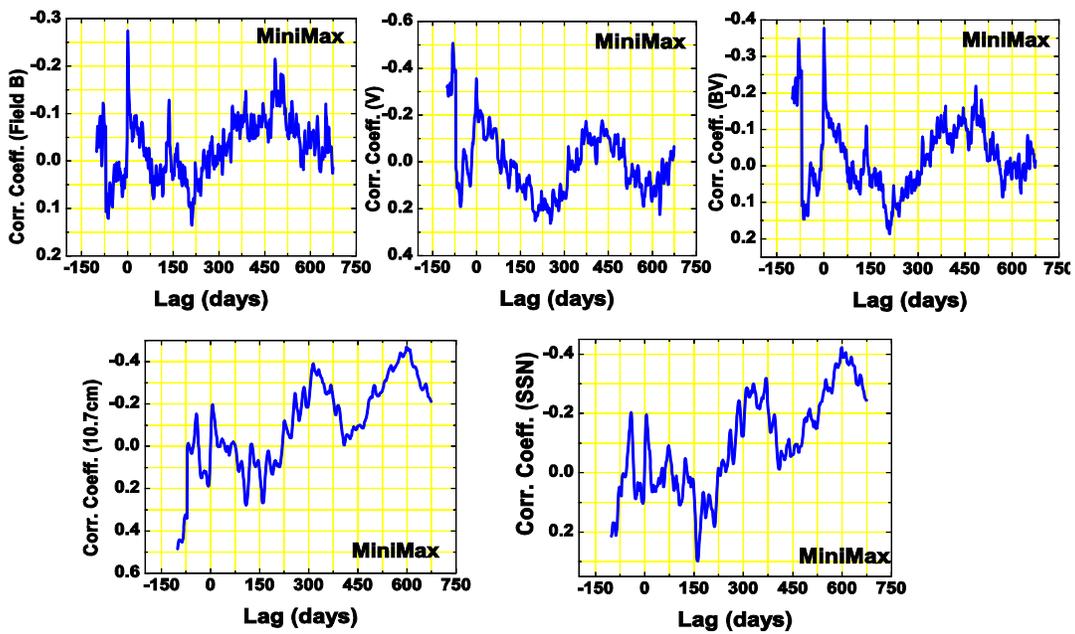

**Figure 7** The variation of time-lag correlation between the daily GCR intensity and solar and interplanetary parameters (SSN, 10.7cm solar radio flux, *V*, *B* and *BV*) during the MiniMax phase.

The 27-day periodicity (and possibly its harmonics) in time lag plots of GCR with $V$ and GCR with $B$ is more clearly visible in Figure 8, in addition to the 27-day periodicity in GCR with SSN. Another important point to be noted from these plots is that the 27-day peaks occur 2 to 3 days later in the GCR with $B$ plot and 6 to 8 days later in the GCR with SSN plot as compared to the GCR with $V$ plot (in this later case peaks occur at 0, 27, 54,…. days).

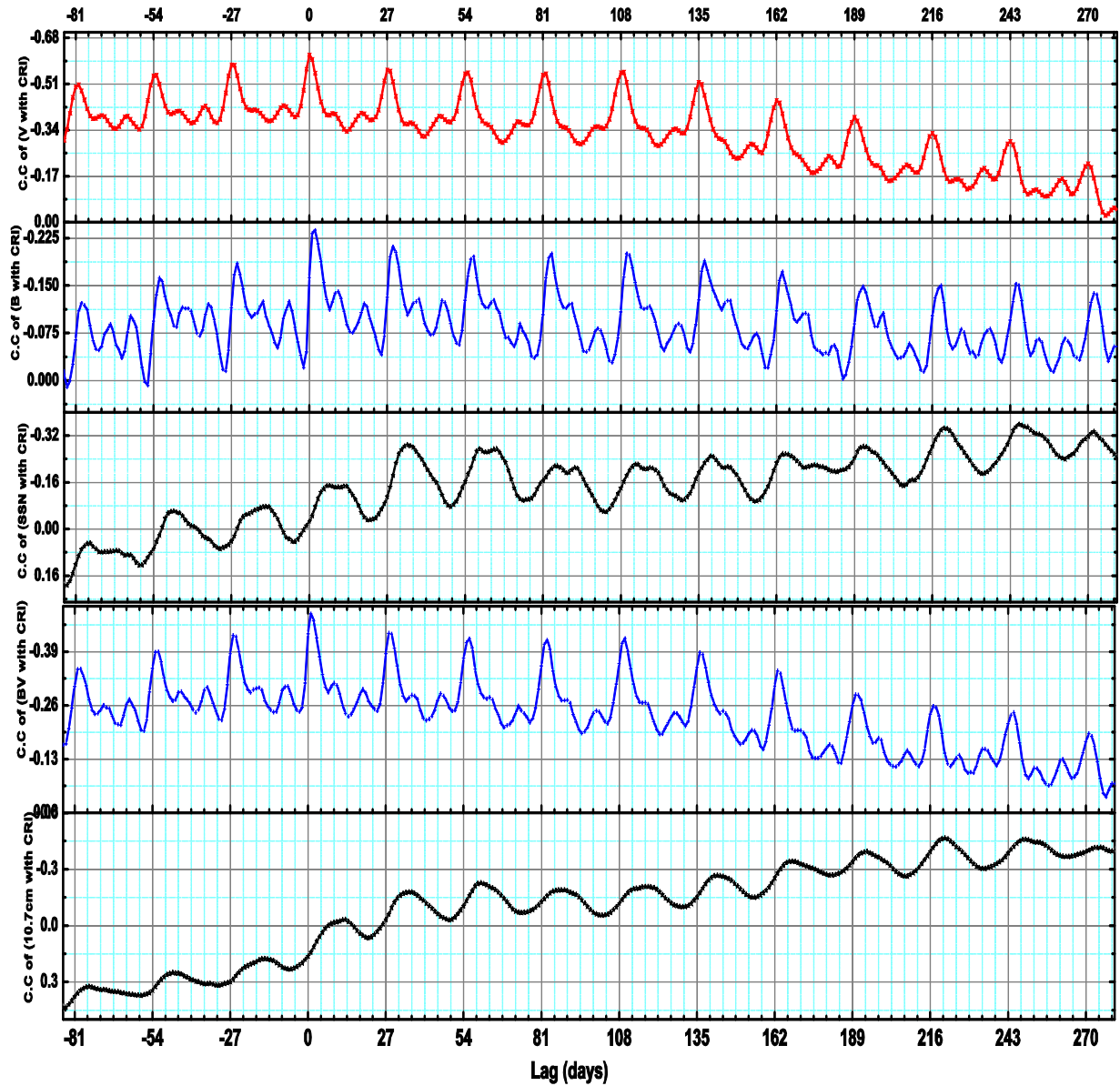

**Figure 8** The periodic variation of the correlation coefficients between GCR intensity and 10.7 cm solar radio flux, $BV$, SSN, $B$ and $V$ during DeepMin with a daily resolution.

2.3 Dependence of GCR Modulation on Solar and Interplanetary Parameters during DeepMin, MiniMax and Ascending Phases of Solar Cycle 24

The transport of cosmic rays in the heliosphere is described by Parker's (1965) transport equation,

$$\frac{\partial f}{\partial t} = -\boldsymbol{V} \cdot \nabla f - \boldsymbol{V}_d \cdot \nabla f + \nabla \cdot (K_s \cdot \nabla f) + \frac{1}{3} (\nabla \cdot \boldsymbol{V}) \frac{\partial f}{\partial \ln P}$$

In this expression, the four terms in the right hand side represent each basic mechanism of cosmic-ray modulation, *i.e.*, solar wind convection ($\boldsymbol{V} \cdot \nabla f$), drifts ($\boldsymbol{V}_d \cdot \nabla f$), diffusion ($\nabla \cdot (K \cdot \nabla f)$) and adiabatic energy change ($\frac{1}{3} (\nabla \cdot \boldsymbol{V}) \frac{\partial f}{\partial \ln P}$). Where $f$ is the cosmic ray distribution function, $V$ is solar wind velocity, $P$ is rigidity, $t$ is time, $V_d$ is particle drift velocity, and $K_s$ is the symmetrical diffusion tensor.

In the above transport equation outward convection by the solar wind velocity ($V$), adiabatic energy change depending on the sign of the divergence of $V$, diffusion caused by turbulent irregularities in the heliospheric magnetic field ($B$) gradient, curvature and current sheet drifts sensitive to the tilt of the current sheet (  ) in the global heliospheric magnetic field, depend directly or indirectly on the heliospheric parameters $V$, $B$ and    (see Heber, 2013; Potgieter, 2014). The GCR intensity varies in antiphase with solar activity parameters, *e.g.*, sunspot number and 10.7 cm solar radio flux. Thus, the study of the relationship between changes in GCR intensity and various solar (SSN and 10.7 cm solar radio flux) and heliospheric parameters ($V$, $B$,   ), especially the time delay (shorter or larger), dependence (stronger or weaker) and variation (faster or slower) between GCR intensity and various solar and heliospheric parameters (SSN, 10.7 cm solar radio flux, $V$, $B$, and   ) during different phases of solar cycle is expected to provide indication about their relative importance and, consequently, about physical process involved and their relative contribution.

The optimum value of the linear correlation coefficient obtained during the lag-time calculation of different solar and interplanetary parameters with GCR intensity observed at different ground based NMs is tabulated in Table 2. Considering the values of the coefficients ($r$) as a measure of stronger (high $r$) or weaker (lower $r$) dependence of the GCR intensity modulation on the parameters, we observe that it shows a stronger dependence on solar variability (SSN and 10.7 cm solar radio flux) during the increasing phase of Solar Cycle 24, as compared to DeepMin and MiniMax. Similarly, the GCR modulation shows a stronger dependence on the interplanetary magnetic field ($B$) during the increasing phase of Solar Cycle 24. However, the dependence of GCR intensity modulation on the HCS tilt angle is stronger during both DeepMin and increasing phases as compared to MiniMax. But, it shows a stronger dependence on the solar wind velocity during DeepMin as compared to MiniMax and increasing phase. Although these results are confirmed using a number of NM data with different cutoff rigidities (Table 2), their interpretation requires the knowledge of the level of magnetic field turbulence, as diffusion and drift coefficients are sensitive to this level, which varies with the solar cycle.

Strauss and Potgieter (2014) examined the drift effect during 2009 included in DeepMin when the GCR intensity was at its record high level. They noted that even if the drift velocity ($\boldsymbol{V}_d$) was higher in 2009, the gradient in $f$ ($\nabla f$) was lower caused by the much lower diffusion coefficient, so that the net drift effects ($V_d . \nabla f$) were weaker. We see that the $V$ was very low in 2009 and since $\nabla f$ was also lower in this period, hence the net outward convection ($V . \nabla f$) caused by the solar wind is likely to be much reduced, possibly an additional effect responsible for the record high GCR intensity during deep minimum. This probably indicates that reduced outward convection by the solar wind was one of the additional cause in addition to diffusion and drifts setting up the record high GCR intensity during the DeepMin phase (see also Aslam and Badruddin, 2012).

### 2.4 Energy Dependence of the Rate of Change in GCR Intensity with the Change in Solar and Interplanetary Parameters during DeepMin, MiniMax and Ascending Phases of Solar Cycle 24

The quantitative estimates of the rate of change in GCR intensity recorded at different NMs with different solar and interplanetary parameters have been evaluated using a linear regression analysis after introducing the respective time lags. The slopes of the linear fits between rotation-averaged relative GCR intensity recorded at different NMs and corresponding solar and interplanetary parameters ($I/P$) obtained during the three different phases are tabulated in Table 3a and 3b; here $P$ stands for each analysed parameter.

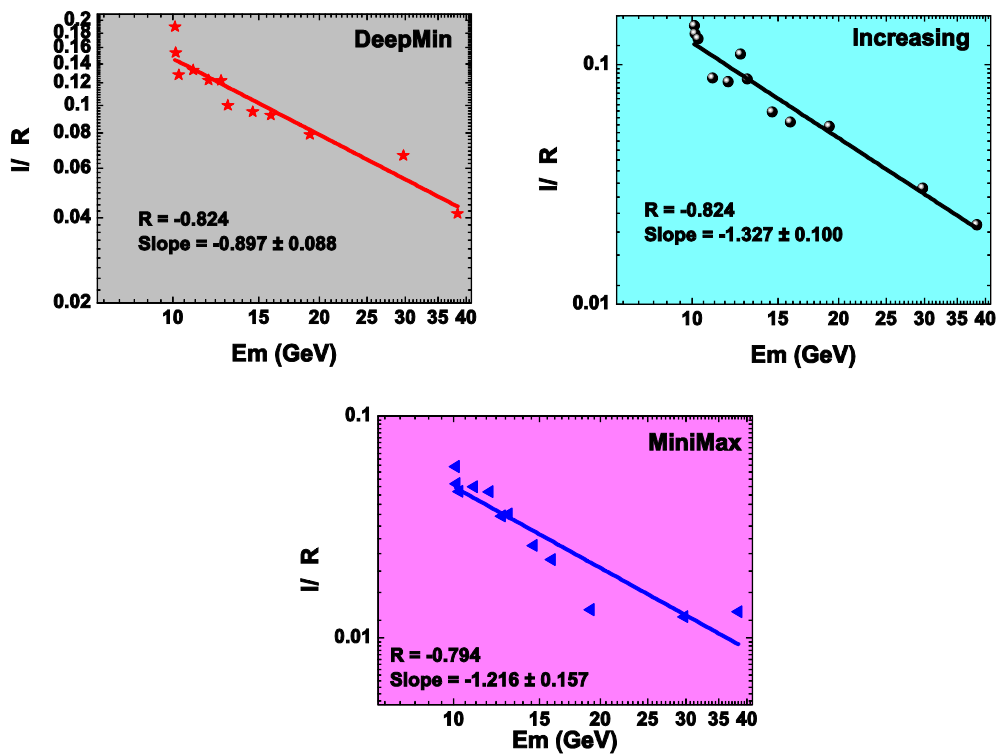

**Figure 9** The variation of the rate of GCR intensity change and change in sunspot number ($I/R$) with the median energy of different stations.

The ratios of rates at which GCR intensity changes with different solar and interplanetary parameters during two different phases is also calculated and the values of the ratios are shown in Table 4a and 4b. This table indicates how fast is the rate of change of GCR intensity during one phase as compared to others. From Tables 3a, 3b, 4a and 4b, we conclude that the GCR intensity decreases at a much faster rate with the change in SSN and 10.7 cm solar radio flux during DeepMin than during MiniMax, whilethis decrease rate is almost same during DeepMin and MiniMax in relation to the HCS tilt angle. However, these decrease rate values during DeepMin and MiniMax are slightly higher than during the increasing phase. As regards the interplanetary parameters, we observe that the GCR intensity decrease with an IMF increase is faster during DeepMin than during MiniMax, while the decrease rate is almost the same in relation to the change in the solar wind velocity during DeepMin and MiniMax. However, the decrease rate in GCR intensity with the change in $V$ is comparatively faster during the increasing phase.

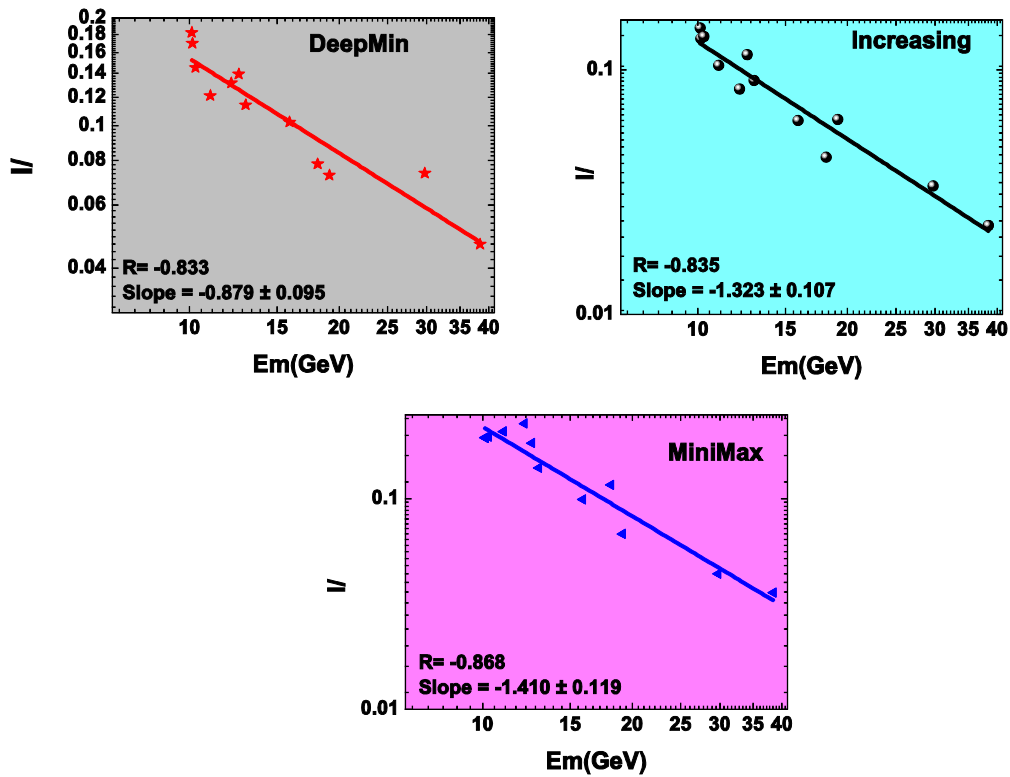

**Figure 10** The variation of the rate of GCR intensity change and change in the HCS tilt ( $I/$ ) with the median energy of different stations.

We have also studied whether the rate of change in GCR intensity with different parameters ( $I/P$ ) shows a similar (or different) dependence on the energy of the particles during the three phases. After calculating the values of median energy of response ($E$m) of different NMs (see Usoskin *et al.*, 2008), we have plotted $I/P$ versus$E$m during the different phases. These plots are shown in Figures 9 – 12. From these plots, we observe that in relation to SSN, the rate of change ( $I/R$ ) is faster with $E$m increase during the increasing and MiniMax phases as compared to the DeepMin phase. However, the tilt angle rate of change ( $I/$ ) with increase $E$m is slower during DeepMin as compared to MiniMax and increasing phases.

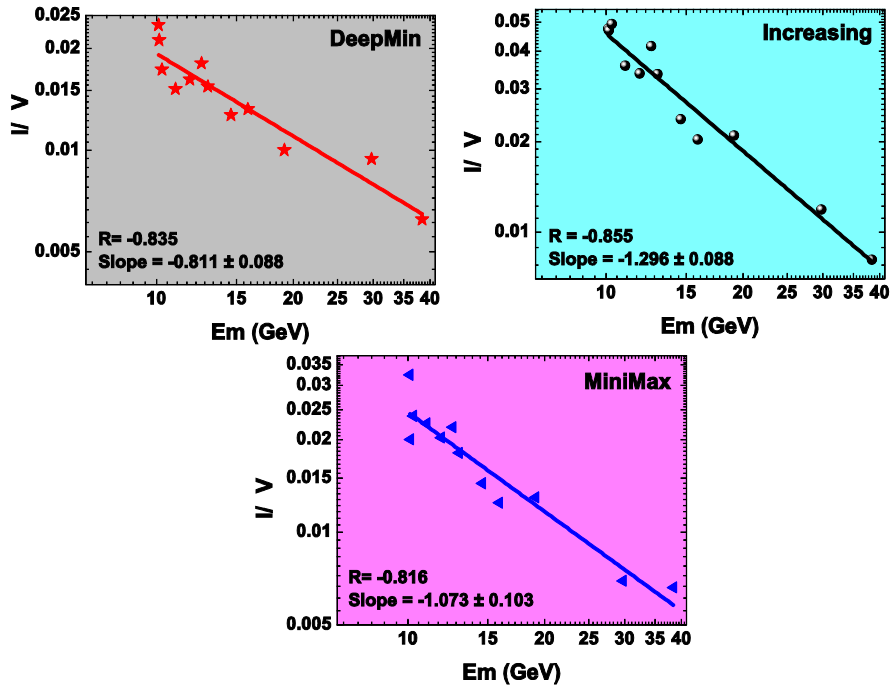

**Figure 11** The variation of the rate of GCR intensity change and change in solar wind velocity ($I/V$) with the median energy of different stations.

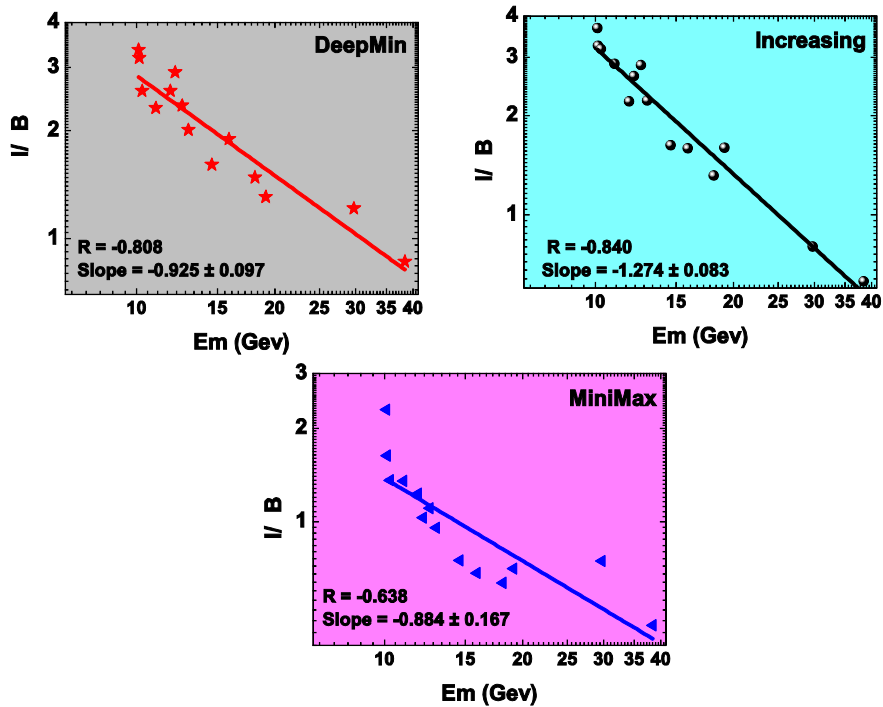

**Figure 12** The variation of the rate of GCR intensity change and change in IMF strength ($I/B$) with the median energy of different stations.

In addition, the $Em$ dependence with $I/V$ and $I/B$ are not equal during the three phases; these rate of changes are larger during the increasing phase as compared with the other two. Although, a linear fit of $I/B$ versus $Em$ plot during MiniMax is also shown, there is large scattering in the data points. A linear curve is expected during this phase; however, the presence of transient fluctuations in GCR intensity and magnetic field data during MiniMax can be responsible for the deviation from linearity.

### 2.5 Energy Spectrum during DeepMin and MiniMax

Using the relative change in GCR intensity with $Em$ during DeepMin and MiniMax, we have studied the energy spectrum during these two peculiar periods (Figure 13). From the spectrum plotted using neutron monitor measurements from pole to equator, we observe almost similar energy spectra during MiniMax (exponent = -1.27) and DeepMin (exponent = -1.37).

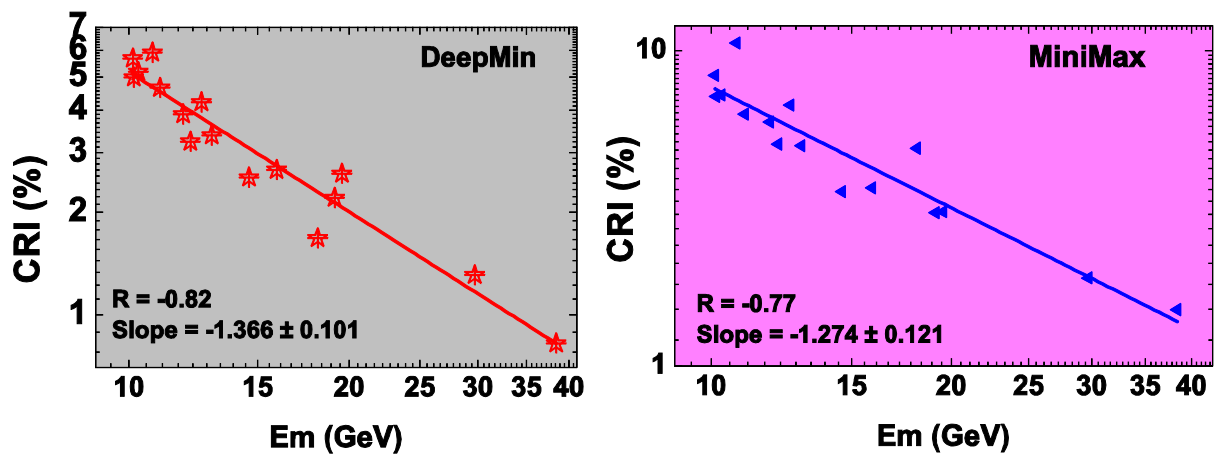

**Figure 13** The energy spectrum during the DeepMin between Solar Cycles 23 and 24 and the MiniMax of Solar Cycle 24.

## 3 Summary of Results and Conclusions

- The time lag between GCR intensity and solar parameters (SSN and 10.7 cm solar radio flux) is not same during the three studied phases; it is shorter (3 solar rotations) during the increasing, intermediate (12 solar rotations) during MiniMax, and longer ( b22 solar rotations) during DeepMin phases.

- Modulation of the GCR intensity shows a stronger dependence on solar variability (SSN, 10.7cm) during the increasing phase of Solar Cycle 24, and a stronger dependence on solar wind velocity during DeepMin between Solar Cycles 23 and 24.

- The change in GCR intensity with the variation of solar and interplanetary, plasma and field, parameters and tilt angle is faster at lower cutoff rigidity stations during all the three phases of solar activity.

- The rate of GCR intensity decrease with IMF strength is faster during the DeepMin than during the MiniMax, while this rate is almost the same in relation to the change in $V$ during DeepMin and MiniMax.

- The GCR intensity decreases at a much faster rate with the change in SSN and 10.7cm flux during DeepMin than during MiniMax.

- The rate of GCR intensity decrease with tilt angle in almost same during DeepMin and MiniMax.

- The energy spectrum in MiniMax is almost similar to that during DeepMin with nearly equal values of the spectral exponent.

- The stronger or weaker relations, faster or slower changes, different energy dependence in the rate of change with different solar and interplanetary parameters, during different phases, suggest that the interplay among the four basic transport mechanisms changes over the solar cycle during DeepMin, MiniMax and increasing phase of Solar Cycle 24.

**Acknowledgments** We thank the station managers of the neutron monitors whose data have been used in this study. Availability and use of solar and plasma and field data from the NASA/GSFC OMNI Web interface and the HCS inclination data the Wilcox Solar observatory, Stanford, are also acknowledged. We also thank the Reviewers and Editor, whose comments and suggestions helped us to improve the paper.

**Table 1** Time lag (in solar rotations) during the three studied phases for solar and interplanetary parameters calculated for GCR intensity observed at NM stations with different cutoff rigidities ($R_c$) and median energies ($E_m$). MIN, INC and MAX correspond to DeepMin, increasing phase, and MiniMax.

| NM Stations | Rc (GV) | Em (GeV) | SSN MIN | INC | MAX | 10.7cm flux (sfu) MIN | INC | MAX | (degree) MIN | INC | MAX | B (nT) MIN | INC | MAX | V (km s⁻¹) MIN | INC | MAX | BV (mV m⁻¹) MIN | INC | MAX |
|---|---|---|---|---|---|---|---|---|---|---|---|---|---|---|---|---|---|---|---|---|
| McMurdo | 0.01 | 10.122 | 19 | 3 | 12 | 20 | 2 | 12 | 0 | 2 | 12 | 0 | 1 | 0/13 | 0 | 1 | 0 | 0 | 1 | 0/14 |
| Inuvik | 0.18 | 10.151 | 23 | 3 | 12 | 20 | 3 | 12 | 1 | 3 | 12 | 0 | 1 | 0/19 | 0 | 1/3 | 0 | 0 | 1 | 0/16 |
| Oulu | 0.81 | 10.302 | 23 | 3 | 12 | 22 | 3 | 12 | 1 | 3 | 12 | 0/6 | 1 | 0/18 | 0 | 1 | 0 | 0 | 1 | 0/16 |
| Moscow | 2.46 | 11.030 | 22 | 1 | 12 | 22 | 1 | 12 | -1 | 3 | 12 | 0/3 | 1 | 0/18 | 0 | 1/3 | 0 | 0 | 1 | 0/16 |
| Irkutsk | 3.66 | 11.858 | 23 | 1 | 13 | 21 | 3 | 12 | 0 | 2 | 12 | 0 | 0/5 | 0/13 | -1 | 1/3 | 0 | 0 | 1 | 0/16 |
| Jungfraujoch | 4.48 | 12.570 | 24 | 3 | 12 | 21 | 2 | 12 | 1 | 3 | 11 | 0/5 | 1 | 0/13 | 1 | 1 | 0 | 2 | 1 | 0/16 |
| Hermanus | 4.9 | 12.980 | 23 | 3 | 12 | 21 | 3 | 12 | 1 | 3 | 12 | 0/6 | 1 | 0/18 | 2 | 1 | 0 | 2 | 1 | 0/16 |
| Rome | 6.32 | 14.596 | 20 | 3 | 12 | 20 | 3 | 12 | -1 | 3 | 11 | 0/6 | 1 | 1 | 0/3 | 1 | 0 | 0/3 | 1 | 0/16 |
| Potchefstrom | 7.3 | 15.918 | 23 | 3 | 12 | 23 | 2 | 12 | 0 | 2 | 11 | 0 | 1 | 0/13 | 1 | 1 | 0 | 0 | | 0 |
| Athens | 8.72 | 18.131 | 20 | 3 | 14 | 20 | 3 | 22 | 1 | 2 | 11 | 0 | 1 | 0/13 | 1 | -1 | 0 | 0 | 1 | 0/16 |
| Tsumeb | 9.29 | 19.120 | 21 | 2 | 11 | 21 | 2 | 12 | 0 | 2 | 11 | 2 | 1 | 1 | 0 | 1 | | 0 | 1 | 0 |
| Mexico | 9.53 | 19.553 | 23 | 3 | 11 | 20 | 3 | 12 | 2 | 2 | 11 | 0/6 | 1 | 0/18 | 4 | 1 | 0 | 0/5 | 1 | 0/16 |
| Tibet | 14.1 | 29.727 | 23 | 3 | 11 | 22 | 4 | 10 | 1 | 3 | 12 | 0/6 | 1 | 1 | 1 | 1 | 0 | 1 | 1 | 1 |
| THAI | 17.1 | 38.398 | 23 | 3 | 11 | 20 | 3 | 10 | 1 | 3 | 11 | 0 | 1 | 1 | 1 | 1 | 0 | 1 | 1 | 0 |

**Table 2** Optimum values of the correlation coefficient (*r*) between GCR intensity and various parameters during the three studied phases at NM stations with different cutoff rigidities (*R*c) and median energies (*E*m). MIN, INC and MAX correspond to DeepMin, increasing phase, and MiniMax..

| NM Stations | Rc (GV) | Em (GeV) | SSN MIN | SSN INC | SSN MAX | 10.7cm flux (sfu) MIN | 10.7cm flux (sfu) INC | 10.7cm flux (sfu) MAX | (degree) MIN | (degree) INC | (degree) MAX | B (nT) MIN | B (nT) INC | B (nT) MAX | V (km s$^{-1}$) MIN | V (km s$^{-1}$) INC | V (km s$^{-1}$) MAX | BV (mV m$^{-1}$) MIN | BV (mV m$^{-1}$) INC | BV (mV m$^{-1}$) MAX |
|---|---|---|---|---|---|---|---|---|---|---|---|---|---|---|---|---|---|---|---|---|
| McMurdo | 0.01 | 10.122 | -0.739 | -0.883 | -0.501 | -0.776 | -0.874 | -0.557 | -0.922 | -0.923 | -0.76 | -0.627 | -0.755 | -0.579 | -0.903 | -0.692 | -0.565 | -0.893 | -0.845 | -0.573 |
| Inuvik | 0.18 | 10.151 | -0.746 | -0.883 | -0.519 | -0.773 | -0.872 | -0.635 | -0.907 | -0.933 | -0.766 | -0.626 | -0.738 | -0.583 | -0.855 | -0.795 | -0.434 | -0.855 | -0.865 | -0.581 |
| Oulu | 0.81 | 10.302 | -0.756 | -0.868 | -0.488 | -0.797 | -0.851 | -0.591 | -0.929 | -0.944 | -0.776 | -0.625 | -0.746 | -0.551 | -0.851 | -0.8 | -0.524 | -0.845 | -0.875 | -0.576 |
| Moscow | 2.46 | 11.030 | -0.742 | -0.834 | -0.505 | -0.731 | -0.832 | -0.628 | -0.839 | -0.906 | -0.802 | -0.642 | -0.793 | -0.541 | -0.803 | -0.741 | -0.49 | -0.803 | -0.827 | -0.643 |
| Irkutsk | 3.66 | 11.858 | -0.774 | -0.887 | -0.521 | -0.784 | -0.844 | -0.659 | -0.889 | -0.913 | -0.593 | -0.657 | -0.711 | -0.516 | 0.855 | -0.765 | -0.569 | -0.854 | -0.824 | -0.36 |
| Jungfraujoch | 4.48 | 12.570 | -0.746 | -0.864 | -0.445 | -0.767 | -0.851 | -0.578 | -0.917 | -0.925 | -0.727 | -0.632 | -0.769 | -0.415 | -0.88 | -0.778 | -0.567 | -0.875 | -0.88 | -0.569 |
| Hermanus | 4.9 | 12.980 | -0.715 | -0.864 | -0.506 | -0.786 | -0.857 | -0.62 | -0.897 | -0.929 | -0.745 | -0.589 | -0.763 | -0.511 | -0.851 | -0.8 | -0.523 | -0.831 | -0.889 | -0.61 |
| Rome | 6.32 | 14.596 | -0.722 | -0.871 | -0.493 | -0.814 | -0.859 | -0.586 | -0.859 | -0.892 | -0.716 | -0.6 | -0.771 | -0.47 | -0.853 | -0.781 | -0.563 | -0.835 | -0.888 | -0.549 |
| Potchefstrom | 7.3 | 15.918 | -0.751 | -0.852 | -0.478 | -0.779 | -0.851 | -0.606 | -0.909 | -0.919 | -0.692 | -0.623 | -0.812 | -0.434 | -0.87 | -0.72 | -0.546 | -0.864 | -0.882 | -0.477 |
| Athens | 8.72 | 18.131 | -0.72 | -0.487 | -0.662 | -0.797 | -0.445 | -0.73 | -0.837 | -0.534 | -0.705 | -0.598 | -0.539 | -0.53 | -0.854 | --0.61 | -0.187 | -0.833 | -0.607 | -0.457 |
| Tsumeb | 9.29 | 19.120 | -0.724 | -0.828 | -0.305 | -0.765 | -0.838 | -0.362 | -0.875 | -0.908 | -0.556 | -0.615 | -0.817 | -0.512 | -0.912 | -0.744 | -0.651 | -0.888 | -0.893 | -0.631 |
| Mexico | 9.53 | 19.553 | -0.726 | -0.836 | -0.393 | -0.765 | -0.831 | -0.515 | -0.899 | -0.897 | -0.672 | -0.601 | -0.758 | -0.508 | -0.883 | -0.755 | -0.538 | -0.854 | -0.864 | -0.567 |
| Tibet | 14.1 | 29.727 | -0.769 | -0.831 | -0.476 | -0.788 | -0.821 | -0.437 | -0.928 | -0.874 | -0.537 | -0.642 | -0.755 | -0.713 | -0.88 | -0.755 | -0.47 | -0.855 | -0.873 | -0.573 |
| THAI | 17.1 | 38.398 | -0.726 | -0.791 | -0.504 | -0.786 | -0.792 | -0.477 | -0.887 | -0.859 | -0.493 | -0.611 | -0.803 | -0.627 | -0.882 | -0.716 | -0.558 | -0.866 | -0.872 | -0.611 |

**Table 3a** Rate of change of GCR intensity with different parameters ($\Delta I / \Delta P$) during different studied phases at NM stations with different cutoff rigidities ($R$c) and median energies ($E$m). MIN, INC and MAX correspond to DeepMin, increasing phase, and MiniMax.

| NM Stations | Rc (GV) | Em (GeV) | SSN MIN | SSN INC | SSN MAX | 10.7cm (sfu) MIN | 10.7cm (sfu) INC | 10.7cm (sfu) MAX | (degree) MIN | (degree) INC | (degree) MAX |
|---|---|---|---|---|---|---|---|---|---|---|---|
| McMurdo | 0.01 | 10.122 | -0.1893 ± 0.035 | -0.145 ± 0.016 | -0.0591 ± 0.021 | -0.2329 ± 0.0386 | -0.1832 ± 0.0208 | -0.0724 ± 0.0208 | -0.182 ± 0.0156 | -0.148 ± 0.0126 | -0.1948 ± 0.0340 |
| Inuvik | 0.18 | 10.151 | -0.1531 ± 0.027 | -0.1345 ± 0.014 | -0.0494 ± 0.017 | -0.2201 ± 0.0368 | -0.1711 ± 0.0196 | -0.0665 ± 0.0165 | -0.17 ± 0.016 | -0.1339 ± 0.0106 | -0.1935 ± 0.0331 |
| Oulu | 0.81 | 10.302 | -0.1281 ± 0.022 | -0.1279 ± 0.015 | -0.0457 ± 0.017 | -0.1822 ± 0.0309 | -0.1615 ± 0.0204 | -0.0610 ± 0.017 | -0.1452 ± 0.0118 | -0.1366 ± 0.0097 | -0.1965 ± 0.0326 |
| Moscow | 2.46 | 11.030 | -0.1328 ± 0.024 | -0.0878 ± 0.012 | -0.0479 ± 0.017 | -0.1334 ± 0.0356 | -0.1277 ± 0.0203 | -0.0656 ± 0.0166 | -0.1212 ± 0.0161 | -0.1042 ± 0.0099 | -0.2081 ± 0.032 |
| Irkutsk | 3.66 | 11.858 | -0.1226 ± 0.020 | -0.0847 ± 0.009 | -0.0455 ± 0.018 | -0.1735 ± 0.0280 | -0.1306 ± 0.0188 | -0.0664 ± 0.0179 | -0.1276 ± 0.013 | -0.1024 ± 0.0093 | -0.101 ± 0.0323 |
| Jungfraujoch | 4.48 | 12.570 | -0.122 ± 0.022 | -0.1105 ± 0.013 | -0.0354 ± 0.015 | -0.1773 ± 0.0303 | -0.1368 ± 0.0182 | -0.0506 ± 0.0146 | -0.1392 ± 0.012 | -0.1152 ± 0.0097 | -0.1835 ± 0.0354 |
| Hermanus | 4.9 | 12.980 | -0.0997 ± 0.020 | -0.0869 ± 0.010 | -0.0361 ± 0.021 | -0.1531 ± 0.0246 | -0.1109 ± 0.0136 | -0.0486 ± 0.0126 | -0.1142 ± 0.0115 | -0.0904 ± 0.0073 | -0.14 ± 0.0256 |
| Rome | 6.32 | 14.596 | -0.0948 ± 0.019 | -0.0634 ± 0.007 | -0.026 ± 0.0094 | -0.1259 ± 0.0183 | -0.0806 ± 0.0098 | -0.0339 ± 0.10 | -0.0738 ± 0.013 | -0.0627 ± 0.0068 | -0.1172 ± 0.0233 |
| Potchefstrom | 7.3 | 15.918 | -0.0922 ± 0.017 | -0.0576 ± 0.007 | -0.0225 ± 0.0084 | -0.1292 ± 0.0225 | -0.0735 ± 0.0093 | -0.0313 ± 0.0084 | -0.1021 ± 0.0095 | -0.062 ± 0.0054 | -0.0991 ± 0.0211 |
| Athens | 8.72 | 18.131 | -0.0847 ± 0.017 | -0.0428 ± 0.016 | -0.0326 ± 0.0076 | -0.1105 ± 0.0171 | -0.0491 ± 0.0206 | -0.0301 ± 0.0067 | -0.0781 ± 0.0104 | -0.0438 ± 0.014 | -0.1159 ± 0.0271 |
| Tsumeb | 9.29 | 19.120 | -0.0785 ± 0.015 | -0.0551 ± 0.008 | -0.0134 ± 0.0085 | -0.0974 ± 0.0167 | -0.0723 ± 0.0096 | -0.0163 ± 0.0086 | -0.0725 ± 0.0081 | -0.0626 ± 0.006 | -0.0678 ± 0.0207 |
| Mexico | 9.53 | 19.553 | -0.1094 ± 0.021 | -0.0591 ± 0.008 | -0.0217 ± 0.0101 | -0.1600 ± 0.0275 | -0.0756 ± 0.0103 | -0.0299 ± 0.0102 | -0.1348 ± 0.0133 | -0.0625 ± 0.0063 | -0.1065 ± 0.0239 |
| Tibet | 14.1 | 29.727 | -0.0664 ± 0.011 | -0.0305 ± 0.004 | -0.0124 ± 0.0066 | -0.0914 ± 0.0166 | -0.0395 ± 0.0058 | -0.0129 ± 0.0076 | -0.0736 ± 0.006 | -0.0334 ± 0.0038 | -0.044 ± 0.0185 |
| THAI | 17.1 | 38.398 | -0.0413 ± 0.008 | -0.0214 ± 0.003 | -0.0131 ± 0.0046 | -0.0621 ± 0.0010 | -0.0276 ± 0.0044 | -0.0133 ± 0.0050 | -0.0466 ± 0.005 | -0.0231 ± 0.0028 | -0.0358 ± 0.0129 |

**Table 3b** Rate of change of GCR intensity with different parameters ($\Delta I/\Delta P$) during different studied phases at NM stations with different cutoff rigidities ($Rc$) and median energies ($Em$). MIN, INC and MAX correspond to DeepMin, increasing phase, and MiniMax.

| NM Stations | Rc (GV) | Em (GeV) | $B$ (nT) | | | $V$ (km s⁻¹) | | | $BV$ (mV m⁻¹) | | |
|---|---|---|---|---|---|---|---|---|---|---|---|
| | | | MIN | INC | MAX | MIN | INC | MAX | MIN | INC | MAX |
| McMurdo | 0.01 | 10.122 | -3.355 ± 0.851 | -3.675 ± 0.613 | -2.290 ± 0.658 | -0.0234 ± 0.0023 | -0.0475 ± 0.010 | -0.0324 ± 0.0097 | -4.1442 ± 0.43 | -6.2506 ± 0.8063 | -3.4192 ± 0.9986 |
| Inuvik | 0.18 | 10.151 | -3.178 ± 0.81 | -3.249 ± 0.607 | -1.628 ± 0.463 | -0.0211 ± 0.0026 | -0.0468 ± 0.0073 | -0.02 ± 0.0085 | -3.7622 ± 0.4668 | -5.9370 ± 0.7033 | -2.7629 ± 0.7903 |
| Oulu | 0.81 | 10.302 | -2.574 ± 0.68 | -3.176 ± 0.58 | -1.358 ± 0.42 | -0.0173 ± 0.0022 | -0.0493 ± 0.0075 | -0.02387 ± 0.0079 | -3.1320 ± 0.4340 | -5.8093 ± 0.6557 | -2.6991 ± 0.7824 |
| Moscow | 2.46 | 11.030 | -2.310 ± 0.64 | -2.867 ± 0.449 | -1.350 ± 0.428 | -0.0152 ± 0.0023 | -0.0358 ± 0.0063 | -0.02255 ± 0.0082 | -2.5936 ± 0.5283 | -4.6593 ± 0.6458 | -3.0487 ± 0.7418 |
| Irkutsk | 3.66 | 11.858 | -2.575 ± 0.60 | -2.206 ± 0.455 | -1.231 ± 0.48 | -0.0161 ± 0.002 | -0.0337 ± 0.0059 | -0.0203 ± 0.0069 | -2.9038 ± 0.36 | -4.3268 ± 0.6206 | -1.2350 ± 0.755 |
| Jungfraujoch | 4.48 | 12.570 | -2.347 ± 0.69 | -2.841 ± 0.482 | -1.105 ± 0.494 | -0.0180 ± 0.0020 | -0.0415 ± 0.0068 | -0.0219 ± 0.0065 | -3.2327 ± 0.365 | -5.0685 ± 0.5573 | -2.2615 ± 0.6674 |
| Hermanus | 4.9 | 12.980 | -2.005 ± 0.574 | -2.218 ± 0.383 | -0.956 ± 0.328 | -0.0154 ± 0.0019 | -0.0336 ± 0.0052 | -0.01807 ± 0.0060 | -2.4428 ± 0.353 | -4.0282 ± 0.4225 | -2.1713 ± 0.5758 |
| Rome | 6.32 | 14.596 | -1.605 ± 0.458 | -1.624 ± 0.273 | -0.749 ± 0.287 | -0.0127 ± 0.0016 | -0.0237 ± 0.0039 | -0.0144 ± 0.0043 | -1.9244 ± 0.29 | -2.9116 ± 0.3084 | -1.4455 ± 0.4490 |
| Potchefstrom | 7.3 | 15.918 | -1.892 ± 0.485 | -1.588 ± 0.233 | -0.682 ± 0.29 | -0.0132 ± 0.0015 | -0.0203 ± 0.004 | -0.0124± 0.0039 | -2.3102 ± 0.28 | -2.6872 ± 0.2926 | -1.0023 ± 0.3771 |
| Athens | 8.72 | 18.131 | -1.479 ± 0.404 | -1.315 ± 0.428 | -1.033 ± 0.337 | -0.0106 ± 0.0013 | -0.0198 ± 0.0069 | -0.00527 ± 0.0057 | -1.7865 ± 0.242 | -2.3957 ± 0.654 | -1.3278 ± 0.5281 |
| Tsumeb | 9.29 | 19.120 | -1.304 ± 0.376 | -1.596 ± 0.230 | -0.636 ± 0.218 | -0.00998 ± 9.4E-4 | -0.02098 ± 0.0039 | -0.01296 ± 0.0031 | -1.7701 ± 0.2047 | -2.7167 ± 0.2801 | -1.1593 ± 0.2908 |
| Mexico | 9.53 | 19.553 | -1.961 ± 0.65 | -1.547 ± 0.272 | -0.706 ± 0.244 | -0.0178 ± 0.0019 | -0.02228 ± 0.0040 | -0.01379 ± 0.0042 | -2.8578 ± 0.413 | -2.7503 ± 0.3264 | -1.4978 ± 0.4445 |
| Tibet | 14.1 | 29.727 | -1.214 ± 0.359 | -0.802 ± 0.142 | -0.746 ± 0.212 | -0.0094 ± 0.0014 | -0.01189 ± 0.0020 | -0.00694 ± 0.0020 | -1.6154 ± 0.202 | -1.4445 ± 0.1646 | -0.8025 ± 0.3315 |
| THAI | 17.1 | 38.398 | -0.862 ± 0.228 | -0.629 ± 0.095 | -0.463 ± 0.118 | -0.0062 ± 6.8E-4 | -0.0081 ± 0.0016 | -0.00661 ± 0.0020 | -1.080 ± 0.127 | -1.0643 ± 0.1218 | -0.6681 ± 0.1766 |

**Table 4a** Ratio of the rate of change of GCR intensity with different parameters ($\delta I/\delta P$) in for the different studied phases at NM stations with different cutoff rigidities ($R$c) and median energies ($E$m). Min, Inc and Max correspond to DeepMin, increasing phase, and MiniMax.

| NM Stations | Rc (GV) | Em (GeV) | SSN | | | 10.7cm flux (sfu) | | | (degree) | | |
|---|---|---|---|---|---|---|---|---|---|---|---|
| | | | Min/Inc | Min/Max | Inc/Max | Min/Inc | Min/Max | Inc/Max | Min/Inc | Max/Min | Max/Inc |
| McMurdo | 0.01 | 10.122 | 1.306 | 3.203 | 2.453 | 1.271 | 3.217 | 2.53 | 1.23 | 1.07 | 1.316 |
| Inuvik | 0.18 | 10.151 | 1.138 | 3.099 | 2.723 | 1.286 | 3.31 | 2.573 | 1.27 | 1.138 | 1.445 |
| Oulu | 0.81 | 10.302 | 1.002 | 2.803 | 2.799 | 1.128 | 2.987 | 2.648 | 1.063 | 1.353 | 1.439 |
| Moscow | 2.46 | 11.030 | 1.513 | 2.772 | 1.833 | 1.045 | 2.034 | 1.947 | 1.163 | 1.717 | 1.997 |
| Irkutsk | 3.66 | 11.858 | 1.447 | 2.695 | 1.862 | 1.328 | 2.613 | 1.967 | 1.246 | 0.792 | 0.986 |
| Jungfraujoch | 4.48 | 12.570 | 1.104 | 3.446 | 3.121 | 1.296 | 3.504 | 2.704 | 1.208 | 1.318 | 1.593 |
| Hermanus | 4.9 | 12.980 | 1.147 | 2.762 | 2.407 | 1.381 | 3.15 | 2.282 | 1.263 | 1.226 | 1.549 |
| Rome | 6.32 | 14.596 | 1.495 | 3.646 | 2.438 | 1.562 | 3.714 | 2.378 | 1.177 | 1.588 | 1.869 |
| Potchefstrom | 7.3 | 15.918 | 1.601 | 4.098 | 2.56 | 1.758 | 4.128 | 2.348 | 1.647 | 0.971 | 1.598 |
| Athens | 8.72 | 18.131 | 1.979 | 2.601 | 1.314 | 2.251 | 3.671 | 1.631 | 1.783 | 1.484 | 2.646 |
| Tsumeb | 9.29 | 19.120 | 1.425 | 5.871 | 4.121 | 1.347 | 5.975 | 4.436 | 1.158 | 0.935 | 1.083 |
| Mexico | 9.53 | 19.553 | 1.851 | 5.041 | 2.724 | 2.116 | 5.351 | 2.528 | 2.157 | 0.79 | 1.704 |
| Tibet | 14.1 | 29.727 | 2.177 | 5.346 | 2.456 | 2.314 | 7.085 | 3.062 | 2.204 | 0.598 | 1.317 |
| THAI | 17.1 | 38.398 | 1.93 | 3.153 | 1.634 | 2.25 | 4.669 | 2.075 | 2.017 | 0.768 | 1.55 |
| Mean value of ratios | ---- | ---- | 1.51 | 3.61 | 2.46 | 1.60 | 3.96 | 2.51 | 1.47 | 1.12 | 1.58 |

**Table 4b** Ratio of the rate of change of GCR intensity with different parameters ($\partial I / \partial P$) in for the different studied phases at NM stations with different cutoff rigidities ($Rc$) and median energies ($Em$). Min, Inc and Max correspond to DeepMin, increasing phase, and MiniMax.

| NM Stations | Rc (GV) | Em (GeV) | $B$ (nT) | | | $V$ (km s$^{-1}$) | | | $BV$ (mV m$^{-1}$) | | |
|---|---|---|---|---|---|---|---|---|---|---|---|
| | | | Inc/Min | Min/Max | Inc/Max | Min/Inc | Min/Max | Inc/Max | Min/Inc | Min/Max | Inc/Max |
| McMurdo | 0.01 | 10.122 | 1.095 | 1.465 | 1.605 | 2.03 | 1.385 | 1.466 | 1.508 | 1.212 | 1.828 |
| Inuvik | 0.18 | 10.151 | 1.022 | 1.952 | 1.996 | 2.218 | 0.948 | 2.34 | 1.578 | 1.362 | 2.149 |
| Oulu | 0.81 | 10.302 | 1.234 | 1.895 | 2.339 | 2.85 | 1.38 | 2.065 | 1.855 | 1.16 | 2.152 |
| Moscow | 2.46 | 11.030 | 1.241 | 1.711 | 2.124 | 2.355 | 1.484 | 1.588 | 1.796 | 0.851 | 1.528 |
| Irkutsk | 3.66 | 11.858 | 0.857 | 2.092 | 1.792 | 2.093 | 1.261 | 1.66 | 1.49 | 2.351 | 3.503 |
| Jungfraujoch | 4.48 | 12.570 | 1.21 | 2.124 | 2.571 | 2.306 | 1.217 | 1.895 | 1.568 | 1.429 | 2.241 |
| Hermanus | 4.9 | 12.980 | 1.106 | 2.097 | 2.32 | 2.182 | 1.173 | 1.859 | 1.649 | 1.125 | 1.855 |
| Rome | 6.32 | 14.596 | 1.012 | 2.143 | 2.168 | 1.866 | 1.134 | 1.646 | 1.513 | 1.331 | 2.014 |
| Potchefstrom | 7.3 | 15.918 | 0.839 | 2.774 | 2.328 | 1.538 | 0.942 | 1.632 | 1.163 | 2.305 | 2.681 |
| Athens | 8.72 | 18.131 | 0.889 | 1.432 | 1.273 | 1.868 | 0.497 | 3.757 | 1.341 | 1.345 | 1.804 |
| Tsumeb | 9.29 | 19.120 | 1.224 | 2.05 | 2.509 | 2.102 | 1.299 | 1.619 | 1.535 | 1.527 | 2.343 |
| Mexico | 9.53 | 19.553 | 0.789 | 2.778 | 2.191 | 1.252 | 0.775 | 1.616 | 0.962 | 1.908 | 1.836 |
| Tibet | 14.1 | 29.727 | 0.661 | 1.627 | 1.075 | 1.265 | 0.738 | 1.713 | 0.894 | 2.013 | 1.8 |
| THAI | 17.1 | 38.398 | 0.73 | 1.862 | 1.359 | 1.306 | 1.066 | 1.225 | 0.985 | 1.617 | 1.593 |
| Mean value of ratios | ---- | ---- | 0.99 | 2.00 | 1.98 | 1.95 | 1.09 | 1.86 | 1.42 | 1.54 | 2.09 |